\documentclass{aa}
\usepackage{graphicx}
\usepackage{amssymb}
\usepackage{lscape}
\usepackage{natbib}
\bibpunct{(}{)}{;}{a}{}{,}

\newcommand{\beq}{\begin{equation}}
\newcommand{\eeq}{\end{equation}}
\newcommand{\grsim}{\mathrel{\hbox{\rlap{\hbox{\lower4pt\hbox{$\sim$}}}\hbox{$>$}}}}

\begin{document}
\title{Can stellar winds account for temperature fluctuations \\ in {\sc H~ii} regions?} 
\subtitle{The case of \object{NGC 2363}}
\author{V. Luridiana
        \inst{1,2}
	\and
	M. Cervi\~no
        \inst{3} 
        \and
        L. Binette
        \inst{2}
        }

\institute{European Southern Observatory, Karl-Schwarzschild-Str. 2, 
              D-85748 Garching b. M\"unchen, Germany
          \and
Instituto de Astronom\'\i a, Universidad Nacional Aut\'onoma de
M\'exico, Ap. Postal 70-264, 04510 M\'exico D.F., Mexico
          \and 
Max-Planck-Institut f\"ur extraterrestrische Physik, 
              Gie{\ss}enbachstra{\ss}e, D-85748 Garching b. M\"unchen, Germany
}

\offprints{V. Luridiana,
\email{vale@astroscu.unam.mx}}

\date{Received / Accepted }
\authorrunning{V. Luridiana et al. }
\titlerunning{Temperature fluctuations in \object{NGC 2363}}

\abstract{
We compare the rate of kinetic energy injected by stellar winds 
into the extragalactic {\sc H~ii} region \object{NGC 2363} 
to the luminosity needed to feed the observed temperature
fluctuations.
The kinetic luminosity associated to the winds is estimated
by means of 
two different evolutionary synthesis codes,
one of which takes into account the statistical fluctuations
expected in the Initial Mass Function.
We find that, even in the most favorable conditions considered by our model, 
such luminosity is much smaller than the luminosity needed
to account for the observed temperature fluctuations.
The assumptions underlying our study are emphasized
as possible sources of uncertainty affecting our results.
}

\maketitle
\section{Introduction}

The presence of temperature fluctuations in photoionized regions
has been a matter of debate since the pioneering work by \cite{P67}.
Although little doubt on their relevance in real nebulae
seems possible on observational grounds
\cite[see the review by][]{P95},
their inclusion in a theoretical framework is still controversial,
given both the incompleteness of the present theoretical scenario,
and the technical difficulties implied by their inclusion 
in photoionization models.
Plain photoionization theory predicts temperature fluctuations to be very small,
mainly due to the steep dependence of the cooling rate on temperature,
which implies that their lifetime would be quite short;
yet,  the existence of a mechanism steadily providing the energy required to feed them
is not excluded by this line of reasoning.
The importance of the question can be barely overrated given that,
if temperature fluctuations turn out to be as big as observations suggest, 
the present determinations of chemical abundances should be 
rescaled by as much as +0.5 dex.

For a long time, the only way temperature fluctuations were accounted for in theoretical models
was acknowledging the impossibility to reproduce 
the observed intensity of the most affected lines;
see, e.g., \cite{LPL99} for the case of \object{NGC 2363}, 
and  \cite{SS99} for the case of IZw18.
Recently, \cite{BL00} developed a model to quantify the effect of temperature fluctuations on the
ionization and energy balance of nebulae.  In the present work, we use
their schema to investigate whether the kinetic energy provided by 
stellar winds could feed the temperature fluctuations observed in \object{NGC 2363}.
Throughout the paper, by `stellar winds' we generically refer to all those
phenomena involving the ejection of mass from stars into the interstellar
medium (ISM). Given the age range considered, 
the only contributors to stellar winds actually included in the calculations
are hot stars and Wolf-Rayet stars (WRs).

The structure of the paper is as follows: 
in Sect.~\ref{sec:general} we summarize the relevant properties of \object{NGC 2363}.
Sect.~\ref{sec:hotspot} describes the model used to represent 
temperature fluctuations in nebulae.
In Sect.~\ref{sec:mech_lum} we estimate the kinetic luminosity of the stellar
cluster in \object{NGC 2363}, 
and apply the method of Sect.~\ref{sec:hotspot} 
to determine whether stellar winds could be responsible for
the temperature fluctuations observed in \object{NGC 2363}.
Finally, in Sect.~\ref{sec:discussion} we analyze several possible sources of uncertainty,
and in Sect.~\ref{sec:conclusion} we summarize the main conclusions.

\section{General properties of \object{NGC 2363}}\label{sec:general}

\object{NGC 2363} is a giant extragalactic {\sc H~ii} region, located in the
south-west end of the irregular galaxy \object{NGC 2366}.  
It is one of the
brightest extragalactic {\sc H~ii} regions known, 
and an ideal subject for our study, since plenty of data
are available, both observational and theoretical.
Two distinctive knots can be clearly distinguished in the region:
throughout this paper, we will always refer to the brightest, youngest
knot, often called `knot A'.

\object{NGC 2363} has been observed by
several groups \cite[e.g., ][]{PPTP86,GDal94,ITL97}, and modeled by
\cite{LPL99} and \cite{Dal00} among others,
while the temperature-fluctuation parameter has been determined 
by \cite{GDal94} in both knots.
 
In the present work, we will assume that
both the stellar cluster and the gas nebula of
\object{NGC 2363} are well described by the best-fit model of \cite{LPL99},
which is a spherical, hollow, radiation-bounded nebula,
consisting of two concentric shells of different densities,
ionized by a young cluster undergoing an extended burst of star formation.
The properties of the model relevant for this work
will be described throughout the text.

\section{The energy implications of temperature fluctuations}\label{sec:hotspot}

To quantify the energy injection rate needed to fuel temperature
fluctuations in a nebula, we follow the model proposed by 
\cite{BL00} and \cite{Bal01}.
The first of these two papers contains a detailed description of the method,
while we refer to the second for an alternative representation of the results.
The model depicts temperature fluctuations 
as a collection of hot spots, 
arising above a uniform equilibrium temperature floor, 
and fed by an unknown heating agent.
This representation allows one to compute 
the average thermal properties of the nebula
in the temperature-fluctuation regime,
but without having to compute real localized fluctuations,
as described below.

\subsection{Definition of temperature fluctuations}\label{sec:t2}

The amplitude of temperature fluctuations in a nebula
is measured by the parameter $t^2$, defined as follows \citep{P67}:

\begin{equation}
t^2={{\int_V (T_{\rm e}{\bf(r)}-T_{\rm 0})^2 N_{\rm e}{\bf(r)} N_{\rm i}{\bf(r)}
dV}\over{T_{\rm 0}^2\int_V N_{\rm e}{\bf(r)} N_{\rm i}{\bf(r)} dV }}\label{eq:t2}
\end{equation}

\noindent where $T_{\rm e}{\bf(r)}$ is the local electron temperature,
$N_{\rm e}{\bf(r)}$ and $N_{\rm i}{\bf(r)}$ are the local values of the electron 
and ionic density respectively, $V$ is the observed volume, and
the average temperature $T_{\rm 0}$ is given by:

\begin{equation}
T_{\rm 0}={{\int_V T_{\rm e}{\bf(r)} N_{\rm e}{\bf(r)} N_{\rm i}{\bf(r)} dV}\over{\int_V
N_{\rm e}{\bf(r)} N_{\rm i}{\bf(r)} dV }}.\label{eq:T0}
\end{equation}

Since the temperature of a model {\sc H~ii} region is not spatially constant,
temperature fluctuations appear and evolve following the evolution of the 
ionization field even in the simplest case of a static, pure photoionization nebula;
in particular, \cite{P97} showed that large temperature fluctuations 
arise during the WR phase, as a consequence of the hardening of the spectrum.
Following the definition by \cite{Fe96},
we will refer to such `structural' temperature fluctuations as $t^2_{\rm str}$.
Typical $t^2_{\rm str}$ values for photoionization models of 
chemically and spatially homogeneous model nebulae 
are in the range $0.00\le t^2_{\rm str}\le 0.02$.
The best-fit photoionization model by \cite{LPL99} yields a value
$t^2_{\rm str}\sim 0.009$.

On the observational side, \cite{GDal94} find for \object{NGC 2363} the value $t^2_{\rm obs}=0.064$
by comparing the Paschen temperature, $T_{\rm e}(Pa)$,
to the [{\sc O iii}] temperature, obtained from the $\lambda 4363/\lambda 5007$ ratio.
Even if in this case the formal error on $t^2_{\rm obs}$ is quite large 
(we calculated $\sigma (t^2_{\rm obs})=0.045$, mainly due to the large error on $T_{\rm e}(Pa)$),
the differences between the model predictions and the observations
seem to imply that an extra-heating mechanism, 
other than photoionization, is at work in most objects,
producing an additional $t^2_{\rm extra}$ such that 
$t^2_{\rm obs} = t^2_{\rm str} + t^2_{\rm extra}$.
In the case of \object{NGC 2363}, we find $t^2_{\rm extra} = 0.055\pm 0.045$.
Note that such value still leaves the door open to the possibility, small but not negligible, 
that $t^2_{\rm extra} = 0.00$; 
however, we assume this not to be the case.
In this respect, it is interesting to note that
$t^2_{\rm obs}=0.098$ in knot B of the same region \citep{GDal94}.
Given the presumed age and metallicity of this region \citep{LPL99}, 
and assuming the same filling factor as in knot A, 
we can roughly estimate $t^2_{\rm str} = 0.02$  \citep[see, e.g.,][]{P97},
yielding $t^2_{\rm extra} = t^2_{\rm obs} - t^2_{\rm str} = 0.078$, 
with an associated error of about $\sigma (t^2_{\rm obs})=0.019$.

More generally, even though the errors might be compatible with $t^2_{\rm extra}=0$ 
in individual cases of ionized regions with $t^2_{\rm obs}>0$,
this is certainly not the case when large samples are considered.
As an example, galactic {\sc H~ii} regions and planetary nebulae collectively show higher
values, typically around $0.04$ \citep[e.g.,][ and references
therein]{Pal95,P95}.  

The presence of temperature fluctuations inside \object{NGC 2363} bears 
important cosmological consequences,
as it affects the determination of the chemical abundances,
lowering the extrapolated primordial helium abundance $Y_{\rm p}$
\citep[e.g.,][and references therein]{PPL01a}.
In our case, $t^2_{\rm obs}=0.064$ implies a downward change in the helium abundance 
of \object{NGC 2363} of order 2\%,
the exact figure depending on which He lines the abundance determination is based on
\citep[see also][]{GDal94}.
It is important to note, however, that temperature fluctuations are not the only factor
affecting a proper chemical abundance determination,
other being, e.g, the ionization structure and the collisional
excitation of the hydrogen lines \citep[][]{SI01,PPL01b}.

\subsection{The hot-spot scheme}

The overall energy balance in a nebula 
is described by the following equation:

\begin{equation}
H = G \label{eq:therm_eq}
\end{equation}

\noindent where $H$ and $G$ are the heating and cooling rates respectively.
In a nebula in ionization equilibrium, with no extra-heating sources,
the heating is provided by photoionization, and the cooling takes place
mainly through collisionally excited line emission and free-free radiation.
A local equilibrium temperature $T_{\rm eq}$ is implicitly defined by
Eq. \ref{eq:therm_eq}, such that the heating terms counterbalance the
cooling terms. This generic model is the reference $t^2_{\rm extra}=0$ model.

In the hot-spot model, a particular temperature profile of a hypothetic nebula
is postulated, consisting of a collection of randomly-generated hot spots 
arising above the $T_{\rm eq}$ floor.
Following the definitions of Eqs.~~\ref{eq:t2} and \ref{eq:T0},
the $t^2_{\rm extra}$ and $T_{\rm 0}$ values can be computed,
giving $t^2_{\rm extra}>0$ and $T_{\rm 0}>T_{\rm eq}$.
The recombination rates of this hypothetic model depend on the assumed 
temperature structure, being in general different from those
of the reference  $t^2_{\rm extra}=0$ model.
It is possible to account for such variation by introducing  
a new `global' temperature, $T_{\rm rec}$, such that 
the intensity of a recombination line is given by:

\beq
I_{\rm rec} \propto T_{\rm rec}^\alpha,\label{eq:trec}
\eeq

\noindent where $\alpha$ is a representative average of the {\sc H~I}
recombination exponent $\alpha(T_{\rm e})$ in the appropriate temperature range.

Analogously, the intensity of a collisionally excited line
in the temperature-fluctuation regime can be calculated
by means of a `global' collisional temperature $T_{\rm col}$,
such that the collisional excitation and de-excitation rates 
are proportional to $T_{\rm col}^\beta\, exp\,[-\Delta E_{ij}/kT_{\rm col}]$
and $T_{\rm col}^\beta$ respectively, with the $\beta$ and $\Delta E_{ij}$
values appropriate for each considered line.

Through the definition of these corrected temperatures,
the new line emissivities can be computed in a straightforward way.
In the hot-spot model, the relationship between $t^2_{\rm extra}$ and 
the derived temperatures is put in analytical form and generalized,
so that the effects of fluctuations on the output quantities
can be easily computed.
For each postulated $t^2_{\rm extra}$ value, a plain photoionization model
is calculated, in which the line emissivities are corrected for
the temperature-fluctuation effects,
and the global energy balance is consequently modified.
This energetic change can be expressed
by means of the quantity $\Gamma_{\rm heat}$, defined as follows:

\beq
\Gamma_{\rm heat}=\frac{L_{\rm fluc}-L_{\rm eq}}{L_{\rm eq}+Q_{\rm eq}},
\eeq

\noindent where $L_{\rm fluc}$ and $L_{\rm eq}$ are the energies radiated 
by the nebula through line emission with and without temperature fluctuations
respectively,
and $Q_{\rm eq}$ is the energy radiated through processes other than line emission.
Thus, ${L_{\rm eq}+Q_{\rm eq}}$ is the equilibrium cooling rate, 
and $L_{\rm fluc}-L_{\rm eq}$ gives the extra-luminosity radiated 
because of temperature fluctuations, so that
$\Gamma_{\rm heat}$ gives a measure of 
the excess energy provided by the putative heating mechanism
driving the fluctuations,
and dispersed through collisional-line enhancement.

\begin{figure} 
   \includegraphics{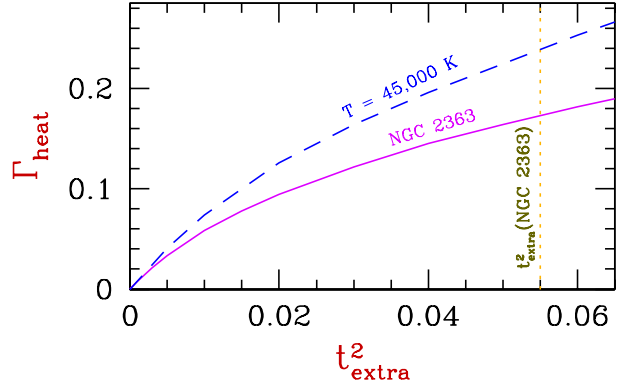}
   \caption{$\Gamma_{\rm heat}$ as a function of $t^2_{\rm extra}$ for the model of \object{NGC 2363} (solid line)
            and for a simpler model (dashed line, see text). The $t^2_{\rm extra}$ value inferred from 
observations of \object{NGC 2363} is also shown.}\label{fig:gamma}
   \end{figure}

The relationship between $t^2_{\rm extra}$ and $\Gamma_{\rm heat}$
depends on the properties of both the ionizing source
and the gaseous nebula.
In Figure~\ref{fig:gamma} we show the dependence of $\Gamma_{\rm heat}$
on $t^2_{\rm extra}$ for the \cite{LPL99} model of \object{NGC 2363}, which we computed 
using a modified version of the photoionization-shock code {MAPPINGS IC}
\citep{Fal97}.
As a general rule,
a tailored model should be computed for each considered case, 
and the calibrations obtained for simpler models
should only be used as rough guidelines when estimating 
the energy implications of temperature fluctuations in a given object.
We can illustrate this point comparing our calibration 
to the one by~\cite{BL00} at a representative temperature.
We first introduce an equivalent effective temperature for the cluster,
defined, according to the method by \cite{MHK91},
as the temperature of an early-type star with the same
{$Q($He$^0$)}/{$Q($H$^0$)} ratio as the synthetic stellar energy distribution (SED).
Using the calibration by \cite{P73}, we found 
an equivalent effective temperature $T_{\rm eff} = 43700\; {\rm K}$.
In Figure~\ref{fig:gamma} we reproduce
the calibration computed by~\cite{BL00} for the case
of a constant-density, Z=0.004 nebula ionized by an unblanketed LTE
atmosphere of T=45,000 K \citep{HM70}. 
Although the two curves show the same qualitative behavior,
they rapidly diverge for increasing $t^2_{\rm extra}$ values.

\section{The kinetic luminosity of the stellar cluster}\label{sec:mech_lum}

In this section, we will compare the kinetic luminosity of \object{NGC 2363}
to the heating rate implied by the observed temperature fluctuations.
To estimate the kinetic luminosity produced by the stellar winds in \object{NGC 2363},
we considered a synthetic stellar population with the parameters resumed in Table~\ref{tab:SED}.
The only difference with respect to the model by \cite{LPL99}
is the adoption of $Z=0.004$ instead of $Z=0.005$,
to avoid interpolation between spectra, and to allow a more direct
comparison with the models by \cite{BL00}.
This choice has little influence on the temperature-fluctuation problem, 
since we estimate that adopting $Z=0.005$ would increase the kinetic luminosity
by less than 50\% \citep[][]{LRD92}, 
which is not sufficient to change, as we will see, the results of this work.

\begin{table}
\begin{tabular}{ll}
Parameter & Value \\
\hline
 SFR                   &   Extended burst, $\Delta t=1.4$\\
 $1+x$                 &   2.00                 \\
 $M_{\rm up}/M_\odot$      &   120                  \\
Age (Myr)              &   3.0                  \\ 
$Z$                    &   0.004                \\
$M_{\rm tot} (1\,\rm{M}_\odot\rightarrow 120\,\rm{M}_\odot)$ & $1.07\times10^5 \;\rm{M}_\odot$\\
\hline
\end{tabular}
\caption{Adopted parameters for the stellar population in \object{NGC 2363}.} \label{tab:SED}
\end{table}

Three different estimates of the kinetic luminosity $L_{\rm kin}$ and the total integrated wind energy for this source are listed
in Table~\ref{Kin_luminosities}.  
One of them was obtained with the first release of {\it Starburst99} \citep[see][]{Lal99} 
run with a theoretical wind treatment,
which is the default set by the code and the one preferred by the
authors of {\it Starburst99}.
The other two were obtained
with an updated version of the synthesis code of \cite{CMH94} (hereafter CMHK, see \cite{Cetal01} for details),
with the wind treatment illustrated in \cite{Cetal01}.
The reasons for considering two different synthesis codes with two different mass-loss laws
reside in the importance of showing the effects of 
different assumptions on the wind treatment
(a possibility granted by the {\it Starburst99} code,
which can be set to any of four different laws),
together with the importance of estimating the expected statistical dispersion
in the energy-related quantitities (a possibility granted by the CMHK code,
either analytically or by means of Monte Carlo simulations).
Stated otherwise, it is not our intention to compare the two codes, but rather the two wind treatments:
had we chosen the same one as \cite{Cetal01} in the {\it Starburst99} settings, we would have found the same result as 
with the CMHK code.

The two estimates of $L_{\rm kin}$ obtained with the CMHK code
have been obtained with an analytical and a Monte Carlo representation of the IMF respectively,
and they both take into account the effect of statistical fluctuations in the IMF.
When statistical effects in the IMF  are taken into account, the
population properties are no more univocally determined, being instead
distributed along a probability distribution curve.
The characteristic parameters of the curve can be determined analytically 
or by means of Monte Carlo simulations \citep[see][]{CVGLMH01}.
The second line of Table~\ref{Kin_luminosities} lists the
analytical estimate of the average wind luminosity and the total integrated wind energy
obtained with the {CMHK} code,
as well as the corresponding 90\% confidence intervals
(i.e., the $5^{\rm th}$ and the $95^{\rm th}$ percentile).
The third line quotes the same quantitites obtained by means of
Monte Carlo simulations,
again with the uncertainties corresponding to the 90\% confidence interval.
These two estimates of $L_{\rm kin}$ are in excellent agreement,
showing the consistency of the two statistical approaches of the CMHK code. 
In all the cases, the extended burst scenario has been represented as the sum of 
individual instantaneous bursts of different ages and roughly equal masses.
In the case of the Monte Carlo method, different simulations, taken 
from an ensemble of 5000 runs of $10^2$ stars each, have been summed up 
until the mass of each individual cluster was reached. 
The total number of independent Monte Carlo simulations 
of the extended burst obtained through this procedure was 64. 

The average estimates obtained with {\sc CMHK} give roughly 70\% higher
luminosity than the {\it Starburst99} value; this is mainly due to the
difference between the wind treatment adopted in the two codes, and is
not directly related to differences in the spectra, which are, for the scopes of the present work,
largely negligible.
The integrated energy value obtained with {\it Starburst99} is higher than the CMHK one,
a somewhat surprising trend given the corresponding luminosity values.
These figures depend on the winds evolution with time:
the wind treatment adopted in the {\it Starburst99} calculation
gives less energetic winds than the one implemented in the CMHK code at earlier ages, 
while for ages greater than about 2.5 Myr the relationship is inverted.

\begin{table}
\begin{tabular}{llll}
Synthesis code & IMF filling  & $L_{\rm kin}$ & $E_{\rm kin}$\\
&  &{($10^{39}$ erg s$^{-1}$)} &{($10^{52}$ erg)} \\
\hline
{\it Starburst99}       & Analytical            & $0.68$                   & 5.0   \\
{\sc CMHK}                 & Analytical     & ${1.14}_{-0.23}^{+0.27}$ & $4.34_{-0.61}^{+0.68}$\\
{\sc CMHK}                 & Monte Carlo    & ${1.16}_{-0.20}^{+0.37}$ & $4.32_{-0.66}^{+0.88}$\\
\hline
\end{tabular}
\caption{Kinetic luminosities and integrated kinetic energies for the three considered SEDs.}\label{Kin_luminosities}
\end{table}

It is now possible to compute the $\Gamma_{\rm heat}$ values
associated with the kinetic energy input,  
according to the general expression:

\beq
\Gamma^{\rm kin}_{\rm heat}=\frac{L_{\rm kin}}{L_{\rm eq}+Q_{\rm eq}},
\eeq

For simplicity, only one value for the equilibrium cooling rate
will be used in the $\Gamma^{\rm kin}_{\rm heat}$ computations,
resulting from a {\sc Cloudy} model \citep{Fe96} 
of the region run with the {\it Starburst99} SED: 

\begin{equation}
L_{\rm eq}+Q_{\rm eq}=G_{\rm eq}^{\rm cooling}=1.46\times10^{41} {\rm erg}\;{\rm s}^{-1},
\end{equation}

\noindent and only the amount of kinetic luminosity deposited according to each SED
will be varied.
This introduces only a minor approximation, since the differences between
synthesis codes, as well as the stochasticity of the IMF, affect the shape
of the spectrum only marginally before the WR phase, which, 
in our extended-burst model accounts only for a minor fraction
of the flux.
This statement is not valid in general, since 
for WR-dominated bursts,
a large dispersion is expected in the ionizing flux,
hence in the cooling rate: see~\cite{CVGLMH01}.
  
The $\Gamma^{\rm kin}_{\rm heat}$ values obtained are listed in
Table~\ref{Gammas}, together with the inferred $t^2_{\rm extra}$
obtained from the hot-spot model of \object{NGC 2363}
(cf. Figure~\ref{fig:gamma});
we will refer to these $t^2_{\rm extra}$ values with the symbol $t^2_{\rm kin}$ 
to emphasize that they are inferred under the assumption
that the temperature fluctuations are driven by 
the stellar-wind kinetic energy. 
All the $t^2_{\rm kin}$ values obtained are extremely small, 
well below the $t^2_{\rm extra}$ inferred from observations.  
However, there are still a few issues to consider before
drawing any conclusion from the comparison between 
the $t^2_{\rm kin}$ values of Table~\ref{Gammas}
and the $t^2_{\rm extra}$ value of \object{NGC 2363} inferred in Sect.~\ref{sec:t2}.

\begin{table}
\begin{tabular}{llcc}
Synthesis code & IMF filling & $\Gamma^{\rm kin}_{\rm heat}$  & $t^2_{\rm kin}\times 10^3$\\  
\hline
{\it Starburst99}       & Analytical           & 0.0047 &                 0.60    \\
{\sc CMHK}          & Analytical           & ${0.0076}_{-0.0015}^{+0.0021}$ & $0.96_{-0.19}^{+0.30}$ \\
{\sc CMHK}          & Monte Carlo          & ${0.0076}_{-0.0011}^{+0.0028}$ & $0.96_{-0.14}^{+0.40}$ \\
\hline
\end{tabular}
\caption{$\Gamma^{\rm kin}_{\rm heat}$ values associated to the wind luminosities, and inferred $t^2_{\rm kin}$ values.}\label{Gammas}
\end{table}

\section{Discussion}\label{sec:discussion}

\subsection{Wind-luminosity thermalization}\label{sec:therm}

An important issue to consider is the thermalization efficiency of the 
wind kinetic energy.  Even assuming that our estimates of the wind
luminosities in \object{NGC 2363} are highly accurate, the derived 
$t^2_{\rm kin}$ values should be corrected downward to take into account
the fraction of the energy injected to the ISM which is
eventually thermalized.  

1-D hydrodynamical models of the bubble around OB associations
indicate that such fraction is about 80\% \citep[][]{P01,PDC01}. 
An estimate of the thermalization efficiency in the case of \object{NGC 2363} 
can be made by comparing the total kinetic energy injected into the region
since the beginning of the starburst to the observed kinetic energy.
\cite{Ral91} observed in \object{NGC 2363} a bubble with a 200 pc diameter,
expanding with a velocity of 45 km s$^{-1}$, and calculated that the total
kinetic energy involved is $E_{\rm kin}^{\rm obs}\sim 2 \times 10^{52}$ ergs,
in agreement with the estimate made by \cite{LPL99}.
\cite{Ral92} found a high-velocity component in the gas, with a kinetic
energy of $E_{\rm kin}^{\rm obs}\sim 10^{53}$ ergs.  \cite{GDal94} confirmed the presence
of this high-velocity gas, and estimated a kinetic energy of $E_{\rm kin}^{\rm obs}\sim 3
\times 10^{52}$ ergs.

These figures should be compared to the estimated values of the integrated wind kinetic energy,
which are listed in Table~\ref{Kin_luminosities} for the three considered cases.
If we add the kinetic energy of the low-velocity bubble observed by
\cite{Ral91} to that of the high-velocity gas observed by
\cite{GDal94}, we find that the efficiency of the
thermalization of wind kinetic energy is 
close to 0, lowering further the computed $t^2_{\rm kin}$ values.

An independent estimate of the efficiency of the thermalization of the
wind energy can be done by considering 
the X-ray component in the spectrum of \object{NGC 2363} detected by \cite{SS98} with {\sl ROSAT}; 
they found $L_x \grsim 6.6 \times 10^{37}$ erg s$^{-1}$ assuming a distance of 3.44 Mpc to the object,
which becomes $L_x \grsim 7.3\times 10^{37}$ erg s$^{-1}$ rescaling to the distance of 3.8 Mpc assumed
by \cite{LPL99}.
Assuming that the origin of the X-ray component is the reprocessing of the wind kinetic energy,
we can infer typical values for the thermalization efficiency of the order $\grsim$ 10\%.

\subsection{Pure photoionization component}

The value of $t^2_{\rm str}$ could be higher if the ionizing spectrum
turned out to be harder than supposed,
as implied by the X-ray component detected by \cite{SS98}; 
in this case, the need for an extra-heating source 
to be added to the photoionization model would be proportionally smaller.
To investigate this possibility, we computed a {\sc Cloudy} model
with the analytical SED of the CMHK code, modified
to account for the transformation of 20\% of the kinetic energy 
into X rays \citep{CMHK01}.
This experimental model is energetically equivalent 
to a hot-spot model in which the extra heating is provided by the wind kinetic luminosity
thermalized with a 20\% efficiency. 
Indeed, we found that for this model $t^2_{\rm str}=0.009$, 
i.e. the same as the \cite{LPL99} to within 0.001,
confirming that the  wind kinetic luminosity is energetically insufficient
to account for the observed temperature fluctuations.

\subsection{Influence of stellar rotation}

As it has been pointed out by \cite{MM00}, rotation
dramatically changes the properties of massive-star models. 
In particular, it increases the mechanical energy released to the ISM 
\citep[see, e.g.,][]{MM01}. Additionally, the winds of rotating massive stars 
can be highly non-spherical, with strong polar or equatorial structures,
depending on the effective temperature and angular velocity of the star
\citep{MD01}. 

Rotation makes it possible to explain in several cases the observed features of individual stars, but, unfortunately,
the evaluation of its effects 
on the integrated spectrum of star forming regions
is not feasible yet.
First, the distribution of angular velocities of the stars in the cluster
should be known.
Second, the distribution of inclination angles should also be known,
since the emitted luminosity depends on the inclination angle.
Third, it would be necessary to establish new homology relations in order to obtain the
isochrones needed to compute the emission spectrum at any given time.

Thus, we can only predict qualitatively that
rotating stars certainly produce a turbulent interstellar medium,
possibly with strong density and temperature inhomogeneities,
due to both the increased wind energy and the anisotropy of their winds.
From the point of view of our study,
the increased energy injected into the medium, 
would translate into a correspondingly higher $\Gamma^{\rm kin}_{\rm heat}$ value,
and the turbulence created in the ISM would possibly increase
the thermalization efficiency of such energy,
in such a way that the `actual' $\Gamma_{\rm heat}$ of the region would approach
its upper limit $\Gamma^{\rm kin}_{\rm heat}$ (see also Sect.~\ref{sec:therm}).

Thus, at a qualitative level, the effect of stellar tracks with rotation
on our temperature-fluctuation model would be to increase
the temperature fluctuation amplitude theoretically achievable
through energy injection by stellar winds,
as compared to non-rotating models.

\subsection{Distance to \object{NGC 2366}}

The distance to \object{NGC 2366} plays several roles
in our analysis.
First, the average properties of a photoionization model 
constrained by observational data 
depend on the assumed distance in complex ways. 
Second, a smaller distance implies a smaller region, hence
a relatively larger statistical dispersion.
Third, the hot-spot model 
is calibrated to a specific model, so that, should the photoionization
model change, the hot-spot model would also change.

In our analysis, based on the photoionization model by \cite{LPL99}, we assumed for the
distance the value 3.8 Mpc determined by \cite{ST76}. 
However, there are a number of more recent distance determinations
indicating smaller distances:
\cite{Tal91} derived a distance of 3.4 Mpc to \object{NGC 2366} through photographic photometry of its brightest stars;
\cite{Aal95} obtained the value of 2.9 Mpc with CCD photometry of its brightest stars;
\cite{Tal95} determined a distance of 3.44 Mpc from Cepheid light curves and colors;
\cite{J98} obtained the value 3.73 Mpc, with an associated error of 0.04 dex,
from a study of the period-luminosity relationship of the red supergiant variables
in NGC 2366. 

To illustrate how our conclusions would change under a different assumption
on the distance, we consider a value of 3.4 Mpc, i.e.
$\sim$10\% less than the assumed distance.
Given the observed H$\beta$ intensity, the smaller distance yields
a 20\% smaller rate of ionizing photons, Q(H$^0$).
Since in this case the region turns
out to be smaller both in linear and in angular dimensions,
more input parameters should be changed,
in order to fulfill the observational constraint
on the observed size of the nebula,
even if the remaining constraints, such as the relevant line ratios, change negligibly.
Though it is beyond the scope of this paper to calculate a full revised photoionization model,
we can expect that a satisfactory fit could be obtained by means
of relatively small adjustments in the density structure
of the nebula surrounding the ionizing cluster,
with the model stellar population rescaled to a total mass
$M'_{\rm tot}=0.80 \times M_{\rm tot}$, and the other
stellar parameters (e.g., SFR, IMF, etc.) left unchanged.
These changes would translate into a $\Gamma^{\rm kin}_{\rm heat}$
smaller by 20\%, with a statistical dispersion proportionally
larger, due to the larger relative weight of the statistical fluctuations 
in the cluster.

The hot-spot model should also be accordingly modified, to take into account the
properties of the revised model. However, we don't expect it too change
too much, since the ionization parameter of the revised model
would be essentially the same as the one of the old model.
Summarizing, we estimate that a downward revision of the assumed distance
would not significantly change our conclusions.

\subsection{Temperature-fluctuation profile}

In the interpretation of our results, it is important to take into account
that they were obtained for a rather specific temperature-fluctuation pattern.
As stated by \cite{BL00}, the model returns the correct results for fluctuations
resembling those depicted in their Figure 1, but for radically different
patterns of hot spots (different in frequency, width, and/or amplitudes)
the model would only provide a first order estimate of 
the relationship between $t^2$ and $\Gamma_{\rm heat}$;
\cite{BL00} estimate that the total uncertainty on $\Gamma_{\rm heat}$
resulting from this approximation is less than 20\%. 

\section{Conclusions}\label{sec:conclusion}

Our results suggest that the kinetic energy provided to the ISM by the
stars through stellar winds cannot account for the
observed temperature fluctuations in \object{NGC 2363}. This result holds
even if a thermalization efficiency of 100\% is assumed; 
however, the comparison of the observed kinetic energy with the 
theoretical estimate of the wind kinetic energy suggest that
such efficiency is rather low.  These results confirm the
conclusion drawn by \cite{Bal01}, and leave the question of the nature
of energy source of temperature fluctuations open.
New insights into the problem could possibly come from
the use of stellar tracks with rotation, and/or the consideration
of a temperature-fluctuation pattern radically different 
from the one used in the present work.

\begin{acknowledgements}
{VL acknowledges the Observatoire Midi-Pyr\'en\'ees de
Toulouse for providing facilities during the first stage of this research. 
The work of LB was supported by the CONACyT grant 32139-E. 
The authors are also grateful to Manuel Peimbert for several excellent suggestions,
and to the referee for critically reading the paper, and making useful comments.
This research has made use of NASA's Astrophysics Data System Abstract Service.}
\end{acknowledgements}

\end{document}